\documentclass[doublecol]{epl2}

\usepackage{graphics}

\title{Random nonlinear layered structures as sources of photon pairs
for quantum-information processing}
\shorttitle{Random nonlinear
layered structures as sources of photon pairs}

\author{Jan Pe\v{r}ina Jr.\inst{1} \and Marco Centini\inst{2} \and Concita Sibilia\inst{2} \and
Mario Bertolotti\inst{2} }

\shortauthor{Jan Pe\v{r}ina Jr. \etal}

\institute{
 \inst{1} Joint Laboratory of Optics of Palack\'{y}
  University and Institute of Physics of Academy of Sciences of the
  Czech Republic, 17. listopadu 50A, 772 07 Olomouc, Czech
  Republic \\
 \inst{2} Dipartimento di Energetica, Universit\`{a} La Sapienza di Roma,
  Via A. Scarpa 16, 00161 Roma, Italy
  }


\abstract{ Random nonlinear layered structures have been found to
be a useful source of photon pairs with perfectly
indistinguishable un-entangled photons emitted into a very narrow
spectral range. Localization of the interacting optical fields
typical for random structures gives relatively high photon-pair
fluxes. Superposing photon-pair emission quantum paths at
different emission angles, several kinds of two-photon states
(including states with coincident frequencies) useful in
quantum-information processing can easily be generated.}

\begin{document}

\maketitle

Since the pioneering generation of photon pairs by Mandel and
coworkers \cite{Hong1967} photon pairs have been generated in many
nonlinear materials using different configurations and geometries.
Their properties are determined by their source. In general photon
pairs can be divided into entangled and un-entangled pairs in a
given degree of freedom. While entangled photon pairs have been
found useful in many physical experiments testing quantum
mechanics, demonstrating quantum teleportation, etc., un-entangled
photon pairs have been found extraordinarily suitable for almost
all quantum-information protocols \cite{URen2003,URen2005}. The
reason is perfect indistinguishability of two photons comprising a
photon pair that is not 'spoilt' by entanglement and that
guarantees perfect visibility in any interferometric setup.
Moreover if pulsed pumping generates simultaneously $ {\cal N} $
photon pairs we have $ 2{\cal N} $ indistinguishable photons
localized in a very sharp time window. This is ideal for
quantum-information processing.

The first attention to un-entangled photon pairs has been
attracted when two-photon states coincident in frequencies
\cite{Giovannetti2002,Giovannetti2002a,Kuzucu2005} were studied.
It has been shown that nonlinear bulk crystals pumped by a
spatially chirped beam
\cite{Torres2005,Torres2005a,Molina-Terriza2005} can serve as a
source of such states. Also assuming non-collinear geometry,
crystal of a given length and pumping with a suitable waist the
theory \cite{Grice2001,Carrasco2006} predicts an un-entangled
state at the output of a nonlinear crystal. Wave-guiding
structures with perpendicular pumping and counter-propagating
signal and idler beams (\cite{DeRossi2002,Booth2002},
\cite{PerinaJr2008} and references therein) also provide
un-entangled states but they suffer from low photon-pair
generation rates \cite{Lanco2006}. Here we offer an alternative
way of generating photon pairs using random nonlinear layered
structures. Easy and fault-tolerant fabrication represents a great
advantage of this source of un-entangled photon pairs that is able
to deliver sufficient photon fluxes. Moreover its integration into
optoelectronic circuits is possible.

\begin{figure}    
 \centerline{\resizebox{0.6\hsize}{!}{\includegraphics{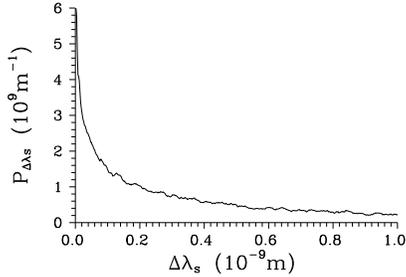}}}
 \caption{Probability distribution $ P_{\Delta\lambda_s} $ of
  widths (FWHM) $ \Delta\lambda_s $ of intensity transmission peaks
  in the signal field at central wavelengths near $ \lambda_0 $;
  $ \theta_s = \theta_i = 10 $~deg.}
 \label{fig1}
\end{figure}

Considering the generation of photon pairs at the fiber-optics
communication wavelength 1.55~$ \mu $m we assume layered
structures made of LiNbO$ _3 $ with etched strips filled by SiO$
_x $ and pumping of spontaneous parametric down-conversion at the
wavelength around $ \lambda_0/2 = 775 $~nm using a pulse 250~fs
long. We note that production of photon pairs in regular layered
structures has been studied in
Refs.~\cite{PerinaJr2006,Centini2005,Vamivakas2004,PerinaJr2007}
where the details of the used model can be found. Structures under
investigation are composed of 300 elementary layers with the mean
layer optical thickness equal to $ \lambda_0/4 $ and material of
each elementary layer is randomly chosen. Typical lengths of these
structures lie around 60~$\mu $m. Numerical simulations for the
used materials have revealed that suitable numbers of layers are
in the interval from 200 to 400. It holds in general that the
greater the contrast of indexes of refraction of two materials the
smaller the number of needed layers. In order to include
fabrication imperfections additional random Gaussian shifts of
boundary positions with variance $ \lambda_0/40 $ are assumed.
Nonlinear material LiNbO$ _3 $ is oriented such that s-polarized
signal, idler, and pump fields can efficiently interact. The
optical axis of LiNbO$ _3 $ is parallel to the planes of
boundaries and coincides with the directions of fields'
polarizations. We note that similar structures have been
investigated from the point of view of second-harmonic generation
\cite{Centini2006,Ochiai2003}. The occurrence of strongly
spatially localized states with highly enhanced electric-field
amplitudes (an optical analog of Anderson localization
\cite{Anderson1958}) is the most striking feature of these
structures. Because the shorter the wavelength the larger the
localization length, structures with localized signal and idler
fields and without localized pump field can be found. In this
case, high values of the overlap integral of fields' amplitudes
inside the structure can be reached. We note that phase-matching
conditions do not play a substantial role here owing to short
lengths of layers. Because frequencies of the signal and idler
fields depend on a given realization of a random structure the
condition for frequency matching can only be fulfilled by an
appropriate choice of the pump-field frequency after fabrication
of the structure. Spectral widths of the signal and idler photons
in a pair are more-less determined by widths of the corresponding
transmission peaks and so their values may vary by several orders
of magnitude, as documented in fig.~\ref{fig1}. This is caused by
the fact that nearly identical conditions for the nonlinear
process are found for all frequencies inside the narrow signal-
and idler-field transmission peaks (provided that the pump-field
spectrum is sufficiently wide). The longer the structure the
narrower (on average) transmission peaks can be expected. Also the
greater the radial emission angle $ \theta $ of a down-converted
field the narrower peaks are found. In a typical sample there
occur several transmission peaks at different radial emission
angles $ \theta $ for a given frequency. Also a given transmission
peak is observed in a certain interval of emission angles $ \theta
$ and it holds that the greater the radial emission angle $ \theta
$ the greater the central frequencies $ \omega_s^0 $ and $
\omega_i^0 $ of the signal and idler fields.

\begin{figure}    
 \centerline{\resizebox{0.8\hsize}{!}{\includegraphics{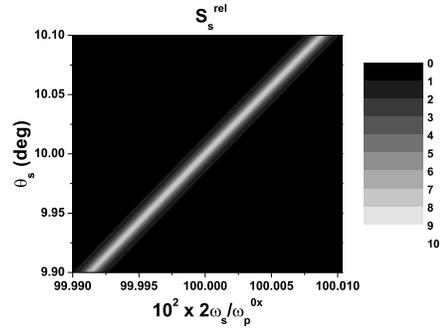}}}
 \caption{Contour plot of relative signal-field energy spectrum $ S_s^{\rm rel} $
 depending on normalized signal-field frequency $ 2\omega_s/\omega_p^{0x} $ and
 signal-field radial emission angle $ \theta_s $. Spectrum  $ S_s^{\rm rel} $ is
 given by the ratio of actual spectrum and that of
 a reference homogeneous structure containing the same amount of
 perfectly-phase-matched nonlinear material.}
\label{fig2}
\end{figure}

Normally incident pump beam in the form of a plane wave and
frequency degenerate signal and idler fields represent the
simplest configuration. In this case, signal and idler photons
occur at opposite sides of the emission cone. Enhancement of
photon-pair generation rate up to three orders of magnitude (for
central frequencies) is observed as a consequence of high signal-
and idler-field amplitudes occurring in localized states (see
fig.~\ref{fig2} where the enhancement one order of magnitude is in
agreement with wider signal- and idler-field spectra). However
high enhancement of photon-pair generation rates is at the expense
of dramatic narrowing of the emission bandwidths and so only from
10 to $ 10^3 $ photon pairs per 100~mW of pumping is expected in
the whole emission cone. The wider the transmission peak the
higher the number of the generated photon pairs. Similar effects
can also be observed in second-harmonic generation
\cite{Centini2006}.

\begin{figure}    
 \centerline{\resizebox{0.8\hsize}{!}{\includegraphics{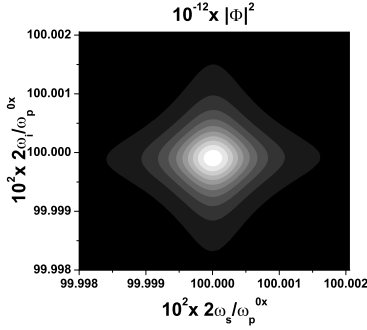}}}
 \caption{Contour plot of squared modulus $ |\phi|^2 $ of the two-photon spectral amplitude
 at signal-field radial emission angle $ \theta_s = 10 $~deg; $ |\phi|^2 $ is normalized such that
 $ 4 \int d\omega_s \int d\omega_i |\phi(\omega_s,\omega_i)|^2
 / (\omega_p^{0x})^2 = 1 $.}
\label{fig3}
\end{figure}

A two-photon state $ |\psi\rangle_{\theta_s,\theta_i} $ describing
a photon pair emitted into the signal- and idler-field radial
emission angles $ \theta_s $ and $ \theta_i $ can be written as
follows:
\begin{equation}
 |\psi\rangle_{\theta_s,\theta_i} = \int d\omega_s \int
  d\omega_i \phi_{\theta_s,\theta_i}(\omega_s,\omega_i)
  |1\rangle_{\theta_s,\omega_s} |1\rangle_{\theta_i,\omega_i},
\end{equation}
where the Fock state $ |1\rangle_{\theta_j,\omega_j} $ contains
one photon at frequency $ \omega_j $ in the form of a plane wave
propagating along radial angle $ \theta_j $ ($ j=s,i $).
Two-photon spectral amplitude $
\phi_{\theta_s,\theta_i}(\omega_s,\omega_i) $ gives us a
probability amplitude of emitting a signal photon at frequency $
\omega_s $ and its idler twin at frequency $ \omega_i $. Its
typical shape with contour plot resembling a cross (see
fig.~\ref{fig3}) is found in random layered structures. This shape
reflects nearly perfect separability of the two-photon state as
can be revealed in the Schmidt decomposition of the two-photon
amplitude $ \phi $. Two photons in a pair are perfectly
indistinguishable owing to identical emission conditions. This can
be experimentally verified, e.g., in Hong-Ou-Mandel interferometer
giving visibility one. Wave-packets with durations in tens or
hundreds of ps characterize these states in time domain.

\begin{figure}    
 a)
 \centerline{\resizebox{0.8\hsize}{!}{\includegraphics{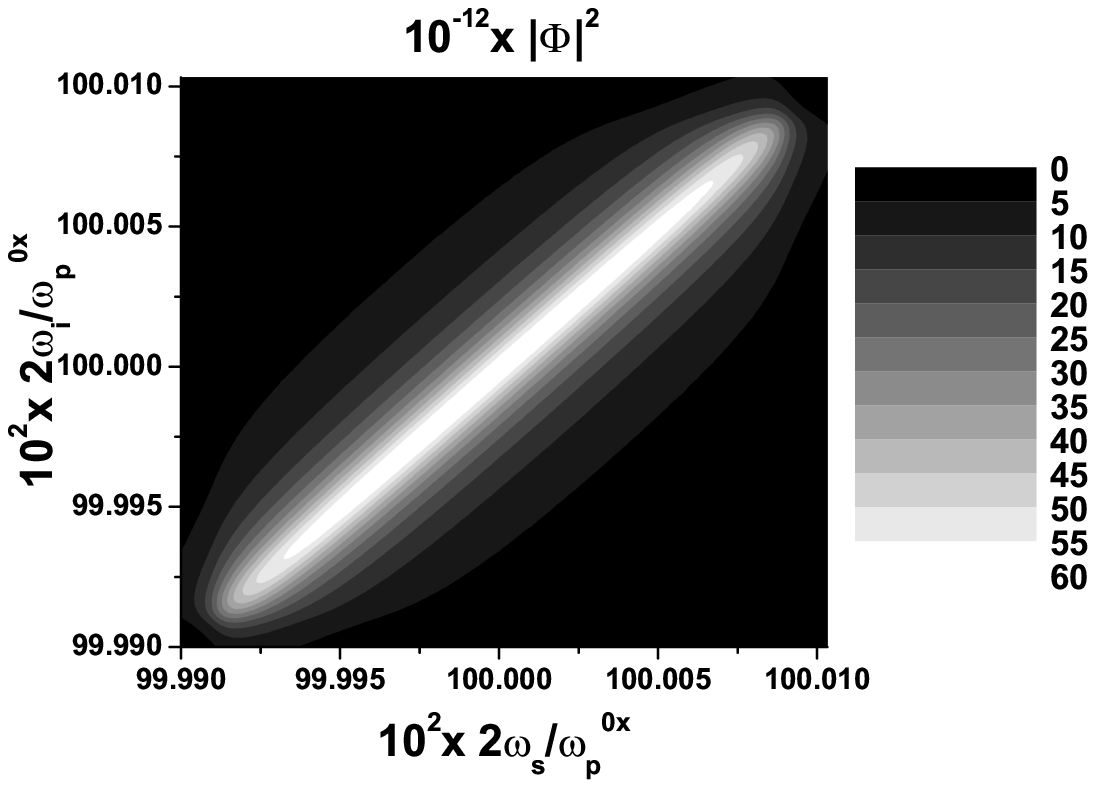}}}

 b)
 \centerline{\resizebox{0.8\hsize}{!}{\includegraphics{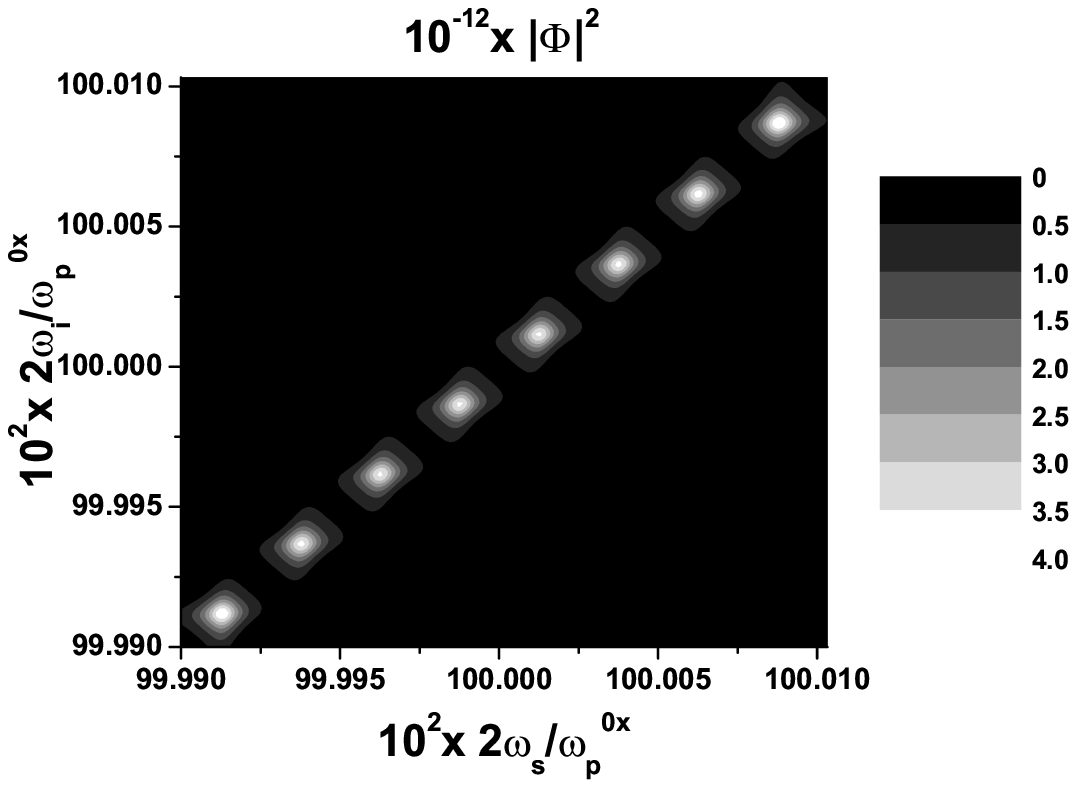}}}
 \caption{Contour plot of squared modulus $ |\phi|^2 $ of the two-photon spectral
 amplitude obtained after superposing two-photon amplitudes from a) a certain range
 of radial emission angles and b) eight equidistantly
 positioned pinholes at increasing radial emission angles; see caption to fig.~\ref{fig3} for
 the normalization of $ \phi $.}
\label{fig4}
\end{figure}

Linear dependence of the signal- and idler-field central
frequencies $ \omega_s^0 $ and $ \omega_i^0 $ on the radial
emission angles $ \theta_s $ and $ \theta_i $ of this source of
un-entangled photon pairs (see fig.~\ref{fig2}) allows to create
easily new two-photon states by superposing photon pairs emitted
into different radial emission angles. The fact that the shapes of
two-photon spectral amplitudes $ \phi $ corresponding to different
emission angles $ \theta_s $ ($ \theta_i $) are nearly identical
is a great advantage and allows to generate well defined
two-photon states. Several kinds of two-photon states needed in
quantum-information protocols can conveniently be generated this
way as the following two examples demonstrate. If two-photon
states emitted into a certain interval of radial emission angles $
\theta $ are superposed (with a suitable phase compensation)
two-photon states with coincident frequencies can be reached (see
fig.~\ref{fig4}a). In this case, the larger the range of the
included emission angles, the broader the intensity spectra of the
signal and idler fields and the higher the entanglement.
Similarly, if two-photon amplitudes propagating from $ M $
equidistantly positioned pinholes at increasing emission angles $
\theta_s $ and $ \theta_i $ are added two-photon states composed
of $ M $ independent collective modes (defined by the Schmidt
decomposition) are obtained (see fig.~\ref{fig4}b).

Random layered structures can also be used for spectrally
non-degenerate emission of photon pairs though the generation of a
suitable structure is an order of magnitude more difficult
compared to the spectrally degenerate case. For example,
un-entangled two-photon states with considerably different signal-
and idler-field spectral widths can be obtained and then exploited
in heralded single-photon sources.

In conclusion, nonlinear random layered structures in which an
optical analog of Anderson localization occurs have been
discovered as sources of photon pairs containing perfectly
indistinguishable photons. Their spectral bandwidths may vary from
0.001~nm up to 1~nm for different realizations of a random
structure. Special two-photon states useful in quantum-information
processing can easily be generated superposing states occurring at
different emission angles.

\acknowledgments
Support by projects IAA100100713 of GA AV \v{C}R,
COST 09026, 1M06002 and MSM6198959213 of the Czech Ministry of
Education as well as support coming from cooperation agreement
between Palack\'{y} University and University La Sapienza in Roma
are acknowledged.

\end{document}